\documentclass[10pt]{iopart}
\usepackage[pdftex]{graphicx}

\newcommand{\bc}{\begin{center}}
\newcommand{\ec}{\end{center}}
\begin{document}

\title[Diagonal and probability representations]
{Superposition rule and entanglement in diagonal
and probability representations of density states\footnote{based
on the invited talk presented by one of us (V.I.M.) at the XV
Central European Workshop on Quantum Optics (Belgrade, Serbia,
30 May -- 3 June 2008).}}

\author{Vladimir I. Man'ko,$^1$ Giuseppe Marmo$^2$ and E. C. George Sudarshan$^3$}

\address{${}^1$ P. N. Lebedev Physical Institute, Leninskii
Prospect 53, Moscow 119991, Russia\\ ${}^2$ Dipartimento di Scienze
Fisiche, Universit\`{a} ``Federico II'' di Napoli\\ and Istituto
Nazionale di Fisica Nucleare, Sezione di Napoli
\\ Complesso Universitario di M. S. Angelo, Via Cintia, I-80126 Napoli,
Italy\\ ${}^3$ Department of Physics,
University of Texas, Austin, Texas 78712, USA}

\ead{mmanko@sci.lebedev.ru}

\begin{abstract}
The quasidistributions corresponding to the diagonal representation of
quantum states are discussed within the framework of operator-symbol
construction. The tomographic-probability distribution describing
the quantum state in the probability representation of quantum mechanics
is reviewed. The connection of the diagonal and probability representations
is discussed. The superposition rule is considered in terms of the
density-operator symbols. The separability and entanglement properties of
multipartite quantum systems are formulated as the properties of the
density-operator symbols of the system states.
\end{abstract}



\section{Introduction}
The pure quantum states are traditionally associated with the wave
function~\cite{Schr26} or a vector in the Hilbert space~\cite{Dirac}. The
mixed quantum states are described by the density matrix~\cite{Landau} or
the density operator~\cite{vonNeumann}. There exist several
representations of quantum states in terms of the quasidistribution
functions like the Wigner function~\cite{Wig32} and the Husimi--Kano
function~\cite{Hus40,Kano56}. The diagonal representation of quantum
states was suggested in \cite{Sud63} (see also \cite{Glauber63}). It
was studied and applied in \cite{Klauder,Mehta}. In this
representation, a quantum state is represented in terms of weighted
sum of coherent-state $|z\rangle$ projectors. The properties of all
the quantum-state representations considered are associated with the
properties of the density operator which is Hermitian, trace-class
nonnegative operator. This means, in particular, that all the
eigenvalues of the density operators must be nonnegative. In the quantum
domain, the multipartite systems have a specific property connected
with strong correlations of the quantum subsystems. This
property provides the entanglement phenomenon~\cite{Schr}.

In the diagonal representation of the density states, the weight function
$\phi(z)$ is an analog of the probability-distribution function in the
phase space. For some class of states, this function is identical to the
probability-distribution function like in classical statistical
mechanics. In \cite{ManciniPLA1996}, the tomographic-probability
representation of quantum states, where the quantum state is associated
with the so-called symplectic tomogram, was introduced. The tomogram
is a fair probability distribution containing the same information
on quantum state that the density operator does (or such its
characteristics as the Wigner or Husimi--Kano functions). The aim of this
work is to find the explicit formulae realizing the connection of the
diagonal and tomographic probability representations. In
\cite{JPAOlgaBeppe}, a review of the star-product-quantization schemes
was given in a unified form. According to this scheme, the functions
like the Wigner function, Husimi--Kano function and
tomographic-probability-distribution function are considered as symbols
of the density operators of a corresponding star-product scheme.
The other goal of our work is to discuss in detail the diagonal
representation within the framework of the star-product scheme along
the lines of construction given in \cite{JPAOlgaBeppe} and to find
mutual relations between the tomographic-probability representation and
the diagonal representation in this context. Using formulation of the
superposition rule in terms of the
density operator~\cite{SudJPA,SudJRLR,SudRev2008}, we consider it within
the framework of the density-state symbols. We focus on the superposition
rule given in terms of tomograms and in terms of weight functions of
the diagonal representation where explicit kernels of the corresponding
star-products are employed to obtain the addition rules for the tomograms
and weight functions. We discuss also the formulation of the separability
and entanglement properties of composed system in the tomographic probability
and diagonal representations.

The paper is organized as follows.

In Section~2, symplectic tomograms and the diagonal representation of
quantum sates are reviewed. In Section~3, the superposition rule
is considered. In Section~4, the diagonal representation and the
star-product formalism are compared. In Section~5, the superposition
rule for tomograms is presented. In Section 6, the entanglement
in the tomographic and diagonal representations is studied. Conclusions
are given in Section~7.

\section{Symplectic tomogram and diagonal representation}
Below we review the approach where the quantum state associated with
tomographic symbol (called symplectic tomogram) of the density operator
(density state) $\hat \rho$ reads (see, for example, \cite{SudRev2008})
\begin{equation}\label{eq.1}
w(X,\mu,\nu)=\mbox{Tr}\,\hat\rho\,\delta(X\hat1-\mu\hat q-\nu\hat
p)\qquad (\hbar=1).
\end{equation}
Here $X,\,\mu,\,\nu\in\mbox{R}$, $\hat q$ and $\hat p$ are the position
and momentum operators, respectively. For the pure state, tomogram is
expressed in terms of the wave function~\cite{Mendes}
\begin{equation}\label{eq.2}
w(X,\mu,\nu)=\frac{1}{2\pi|\nu|}\left|\int\psi(y)\exp\left(\frac{i\mu}{2\nu}y^2
-\frac{iX y}{\nu}\right)\,dy\right|^2.
\end{equation}
The tomogram is nonnegative normalized probability distribution
function of a random variable $X$, i.e.,
\begin{equation}\label{eq.3}
w(X,\mu,\nu)\geq 0,\qquad \int w(X,\mu,\nu) d X=1.
\end{equation}
In the diagonal representation, the density state $\hat \rho$ reads~\cite{Sud63}
\begin{equation}\label{eq.4}
\hat\rho=\int \phi(z)|z\rangle\langle z|\, d\mbox{Re}\,z
\,\,d\mbox{Im}\,z,
\end{equation}
where $|z\rangle=\hat D(z)|0\rangle $ is the coherent state
$\hat a|z\rangle=z|z\rangle$ and the displacement operator
$\hat D(z)=\exp\left(z\hat a^\dagger-z^*\hat a\right)$ is called Weyl system.
Here $\hat a=2^{-1/2}\left(\hat q+i\hat p\right)$ is the boson annihilation
operator and $z$ is a complex number. The probability distribution $w(X,\mu,\nu)$
is expressed in terms of the weight function $\phi(z)$ as follows:
\begin{equation}\label{eq.5}
w(X,\mu,\nu)=\int \phi(z)\langle z|\delta(X\hat 1-\mu\hat q-\nu\hat
p)|z\rangle \,d\mbox{Re}\,z \,\,d\mbox{Im}\,z.
\end{equation}
Using in (\ref{eq.1}) the Fourier decomposition of delta-function and
taking the density state $\hat\rho$ in form (\ref{eq.4}), we obtain for
tomogram
\begin{equation}\label{eq.6}
w(X,\mu,\nu)=\frac{1}{2\pi}\int \phi(z)\langle z|e^{i k(X-\mu\hat
q-\nu\hat p)}|z\rangle\, d\mbox{Re}\,z\,\,d\mbox{Im}\,z,
\end{equation}
where the diagonal matrix element of the operator in the integral can be
considered as the Weyl-system matrix element, i.e.,
\begin{equation}\label{eq.7}
\langle z|e^{- k(i\mu\hat q+i\nu\hat
p)}|z\rangle=\exp\left[z^\ast\alpha-\alpha^\ast
z-\frac{|\alpha|^2}{2}\right],
\end{equation}
with
\begin{equation}\label{eq.8}
\alpha=\frac{k}{\sqrt 2}(\nu-i\mu).
\end{equation}
Evaluating Gaussian integral (\ref{eq.6}), we arrive at
\begin{eqnarray}
\fl w(X,\mu,\nu)=\frac{1}{\sqrt{\pi(\mu^2+\nu^2)}}\nonumber\\
\fl\times\int
\phi(z)\exp\left\{-\frac{\left[X-2^{-1/2}
\Big(z^\ast(\mu+i\nu)+z(\mu-i\nu)\Big)\right]^2}{\mu^2+\nu^2}\right\}\,
d\mbox{Re}\,z\,\, d\mbox{Im}\,z.\label{eq.9}
\end{eqnarray}
The above formula provides the relation of the weight function of the
diagonal representation of the density state and symplectic tomogram of
the quantum state.

For example, the vacuum state $|0\rangle\langle0|$ has the weight
function
\[
\phi_0(z)=\delta\,(\mbox{Re}\,z)\,\,\delta\,(\mbox{Im}\,z).
\]
Formula (\ref{eq.9}) provides tomogram of the ground state
\begin{equation}\label{eq.10}
w_0(X,\mu,\nu)=\frac{1}{\sqrt{\pi(\mu^2+\nu^2)}}\exp\left(-\frac{X^2}{\mu^2+\nu^2}\right).
\end{equation}
This expression can be obtained also by means of formula
(\ref{eq.2}) with
\[
\psi_o(y)=\pi^{-{1/4}}\exp\left(-{y^2}/{2}\right).
\]

\section{Superposition rule for density operators}
For two orthogonal pure states $|\psi_1\rangle$ and $|\psi_2\rangle$,
the superposition rule provides the state
\begin{equation}\label{eq.11}
|\psi\rangle=\sqrt{p_1}\,|\psi_1\rangle+e^{i\phi}\,\sqrt{p_2}\,|\psi_2\rangle,
\end{equation}
which can be realized in the nature as a Schr\"odinger cat state. Here
the positive numbers $p_1$ and $p_2$ satisfy the equality $p_1+p_2=1$
and the phase factor $e^{i\phi}$ determines the interference picture.
The density states $\hat\rho_1=|\psi_1\rangle\langle\psi_1|$ and
$\hat\rho_2=|\psi_2\rangle\langle\psi_2|$ provide the state
$\hat\rho=|\psi\rangle\langle\psi|$, if one uses the nonlinear addition
rule~\cite{SudJPA}
\begin{equation}\label{eq.12}
\hat\rho=p_1\,\hat\rho_1+p_2\,\hat\rho_2+\sqrt{p_1p_2}\,\,\frac{\hat\rho_1\,\hat
P_0\,\hat\rho_2+\hat\rho_2\,\hat
P_0\,\hat\rho_1}{\sqrt{\mbox{Tr}\left(\hat\rho_1\,\hat
P_0\,\hat\rho_2\,\hat P_0\right)}},
\end{equation}
where the operator $\hat P_0$ is a projector ($\mbox{Tr}\,\hat
P_0=1$) which corresponds to the phase term $e^{i\phi}$ in
(\ref{eq.11}).

The superposition rule can be formulated for any symbol of pure density
states $\hat\rho_1$, $\hat\rho_2$ and $\hat\rho$.

\section{Diagonal representation and star-product formalism}
The diagonal representation of density operators can be considered within
the framework of star-product scheme \cite{JPAOlgaBeppe}. Let us
construct two families of operators, which are called dequantizer
\begin{equation}\label{eq.R1}
\hat{\cal U}(z) =\frac{1}{\pi^2}\int
\exp\left(\frac{1}{2}|u|^2+z^\ast u-z u^\ast\right) \hat D(u)\,
d\mbox{Re}\,u\,\, d\mbox{Im}\,u
\end{equation}
and quantizer
\begin{equation}\label{eq.R2}
\hat{\cal D}(z)=|z\rangle\langle z|,
\end{equation}
where $\hat D(u)$ is the Weyl system and $z=x+i y$ is a complex number.
One can check that
\begin{equation}\label{eq.R3}
\mbox{Tr}\, {\widehat{\cal U}}(z)\,{\widehat{\cal
D}}(z')=\delta(x-x')\,\delta(y-y').
\end{equation}
In view of this, one can construct the symbol
of a density operator $\hat\rho$ in the diagonal representation
\begin{equation}\label{eq.R4}
\fl
\phi(z)=\mbox{Tr}\,\widehat{\cal U}(z)\,\widehat\rho
=\frac{1}{\pi^2}\int\exp\left(\frac{1}{2}|u|^2+z^\ast u- z
u^\ast\right)\,\mbox{Tr}\,\widehat\rho\,\widehat D(u)\, d\mbox{Re}\,u
\,\,d\mbox{Im}\, u
\end{equation}
and the reconstruction formula for the density operator reads
\begin{equation}\label{eq.R5}
\widehat\rho=\int\phi(z)\,|z\rangle\langle z| \,d\mbox{Re}\,z\,\,d\mbox{Im}\,z.
\end{equation}
According to \cite{Patr,PatrOlga}, one can construct dual
symbol of the operator $\widehat\rho$
\begin{equation}\label{eq.R6}
\phi^{(d)}(z)=\mbox{Tr}\,\widehat\rho\,|z\rangle\langle z|=\langle
z|\,\widehat\rho|z\rangle
\end{equation}
and dual reconstruction formula
\begin{eqnarray}
\fl\widehat\rho=\int\phi^{(d)}(z)\,\widehat{\cal U}(z)\, d\mbox{Re}\,z\,\,
d\mbox{Im}\,z
\nonumber\\
\fl=\frac{1}{\pi^2}\int\phi^{(d)}(z)\exp\left(\frac{1}{2}|u|^2+z^\ast
u-z u^\ast\right)\widehat D(u)\, d\mbox{Re}\,u \,\,d\mbox{Im}\,
u\,\, d\mbox{Re}\,z \,\,d\mbox{Im}\,z. \label{eq.R7}
\end{eqnarray}
If in (\ref{eq.R4}) the operator $\widehat\rho$ is replaced by some
operator $\widehat A$, the corresponding symbol $\phi_A(z)$ provides the
diagonal representation of the operator. The dual symbol
(\ref{eq.R6}) provides the Husimi--Kano function $Q(z)$. The
reconstruction formula for the density state in terms of the Husimi--Kano
function is just formula (\ref{eq.R7}) with the replacement
$\phi^{(d)}(z)\rightarrow Q(z)$. The duality relation of the diagonal
representation of the density state $\widehat\rho$ and the Husimi--Kano
function was discussed in \cite{Patr,KlaudScag}.

Using the connection of an operator symbol with its
dual~\cite{JPAOlgaBeppe}, one has the connection formula
\begin{equation}\label{eq.R10a}
\fl\phi(z)=\frac{1}{\pi^3}\int
Q(z_1)\exp\left[|u|^2+(z^\ast-z_1^\ast)u-(z-z_1)u^\ast\right]\,
d\mbox{Re}\,u \,\,d\mbox{Im}\, u.
\end{equation}
The inverse formula reads
\begin{equation}\label{eq.R11a}
Q(z)=\int\phi(z) e^{-|z_1-z|^2} \,d\mbox{Re}\,z_1\,\,
d\mbox{Im}\, z_1.
\end{equation}
According to the general formalism~\cite{JPAOlgaBeppe}, the star-product of
symbols related to the diagonal representation is determined by
the kernel
\begin{eqnarray}\label{eq.R11}
\fl K(z_1,z_2,z)=\frac{1}{\pi^2}\mbox{Tr}\left[\int|z_1\rangle\langle
z_1||z_2\rangle\langle z_2|\widehat D(u)\,
\exp\left(\frac{1}{2}|u|^2+z^\ast u-z u^\ast\right)\,
d\mbox{Re}\,u\,\, d\mbox{Im}\, u \right],\nonumber\\
\label{eq.R11}
\end{eqnarray}
which is generalized function of the form
\begin{eqnarray}
\fl K(z_1,z_2,z)=
\frac{1}{\pi^2}\int\exp\,\Big(-(x_2-x_1)^2-(y_2-y_1)^2+(x_2-x_1)a+(y_2-y_1)b\nonumber\\
\fl -i(2y+y_1+y_2)a+i(2x+x_1+x_2)b\Big)da\, \,db,\label{eq.R12}
\end{eqnarray}
where
\[z=x+i y,\quad z_1=x_1+i y_1, \quad z_2=x_2+iy_2.\]

The star-product of symbols of arbitrary operators $\widehat A$ and
$\widehat B$ in the diagonal representation reads
\begin{equation}\label{eq.R13}
\fl (\phi_A\ast\phi_B)(z)=\int K(z_1,z_2,z)\phi_A(z_1)\phi_B(z_2)\,d
\,x_1\,\, d\, y_1\,\, d\, x_2\,\, d\, y_2.
\end{equation}
For example, for the vacuum-state projector $\widehat\rho_0=|0\rangle\langle
0|$ with the weight function -- symbol $\phi_0(z)=\delta(z)$, one has
\begin{equation}\label{eq.R14}
\fl (\phi_0\ast\phi_0)(z)=\int\delta(z_1)\delta(z_2)K(z_1,z_2,z)\, d
\,x_1\,\, d\, y_1\,\, d\, x _2\,\, d \,y_2=\,\delta(z),
\end{equation}
which is equal to $\phi_0(z)$ and corresponds to the pure-vacuum-state
property $\widehat\rho^2_0=\widehat\rho_0$. Tomogram $w(X,\mu,\nu)$ of
the density state $\widehat\rho$ provides the following formula for
the diagonal representation of the density operator
\begin{eqnarray}\label{eq.R15}
\fl\phi(z)=\frac{1}{2\pi^2}\int w(X,\mu,\nu)\exp\left[i
X+\frac{\mu^2+\nu^2}{4}+\frac{z(\nu+i\mu)}{\sqrt2}-\frac{z^\ast(\nu-i\mu)}{\sqrt2}\right]\,
dX\,\, d\mu\,\, d\nu.\nonumber\\
\end{eqnarray}
For example, the vacuum-state tomogram
\[
w_0(X,\mu,0)=\frac{1}{\pi(\mu^2+\nu^2)}\exp\left(-\frac{x^2}{\mu^2+\nu^2}\right)
\]
provides, by means of the above formula, the symbol of the state in the
diagonal representation, i.e., $\delta(z )$.

\section{Superposition rule for tomograms}
The superposition of two pure states with their symbols $\phi_1(z)$
and $\phi_2(z)$ is described by the function
\begin{equation}\label{eq.R16}
\fl \phi(z)=p_1\,\phi_1(z)+p_2\,\phi_2(z)+\sqrt{p_1p_2}\,\,
\frac{(\phi_1\ast\phi_0\ast\phi_2)(z)+(\phi_2\ast\phi_0\ast\phi_1)(z)
}{\sqrt{\int (\phi_1\ast\phi_0\ast\phi_2\ast\phi_0)(z)\,dx\,dy\,}}.
\end{equation}
The star-product in (\ref{eq.R16}) is determined by the kernel (\ref{eq.R12}).

The result obtained can be repeated also for tomographic symbols of the density states.
Thus, the addition rule of two tomographic probabilities of two pure states
$|\psi_1\rangle$ and $|\psi_2\rangle$ reads
\begin{eqnarray}
\fl w(X,\mu,\nu)=p_1\,w_1(X,\mu,\nu)+p_2\,w_2(X,\mu,\nu)\nonumber\\
\fl +\sqrt{p_1p_2}\,\,\frac{(w_1\ast
w_0\ast w_2)(X,\mu,\nu)+(w_2\ast w_0\ast w_1)(X,\mu,\nu)
}{\sqrt{\int \delta(\mu)\,\delta(\nu)\,d\mu\,d\nu\,d\mu'\,d\nu'\,\,\,
\int e^{iX}(w_1\ast w_0\ast w_2\ast w_0)(X,\mu,\nu)\,dX \,}}\,. \label{eq.R17}
\end{eqnarray}
The kernel of tomographic star-product is given in \cite{Patr,PatrOlga}.
The order of integration in the denominator term is essential to obtain the
correct result. In (\ref{eq.R16}) and (\ref{eq.R17}), $\phi_0$ and
$w_0$ are the corresponding symbols of projector $\widehat P_0$ which
determines the relative phase of states $|\psi_1\rangle$ and
$|\psi_2\rangle$ in their superposition.

\section{Entanglement in the diagonal and tomographic-probability representations}
Given bipartite system of a two-mode field.

The tomographic probability distribution is determined as follows:
\begin{eqnarray}\label{eq.R18}
\fl
w(X_1,\mu_1,\nu_1,X_2,\mu_2,\nu_2)=\mbox{Tr}\,\Big[\,\widehat\rho(1,2)\,\delta(X_1\widehat
1-\mu_1\widehat q_1-\nu_1\widehat p_1)\,\delta(X_2\widehat 1-\mu_2\widehat
q_2-\nu_2\widehat p_2)\Big].\nonumber\\
\end{eqnarray}
The density matrix in the diagonal representation is determined by the
symbol of the density state $\widehat\rho(1,2)$
\begin{eqnarray}
\fl\phi(z_1,z_2)=\frac{1}{\pi^4}\int\exp\left[\frac{1}{2}\left(|u_1|^2+|u_2|^2\right)
+z_1^\ast u_1-z_1 u_1^\ast+z_2^\ast u_2-z_2 u_2^\ast \right]\nonumber\\
\fl\times\mbox{Tr}\,\widehat\rho(1,2)\,\widehat D_1(u_1)\,\widehat
D_2(u_2)\,d\mbox{Re}\,u_1\,d\,\mbox{Im}\, u_1
\,d\mbox{Re}\,u_2\,d\mbox{Im}\, u_2.\label{eq.R19}
\end{eqnarray}
The state $\widehat\rho(1,2)$ is separable, if the density state can be
written as a convex sum
\begin{equation}\label{eq.R20}
\widehat\rho(1,2)=\sum_k \,P_k\,\widehat\rho_k(1)\otimes\widehat\rho_k(2),\quad
P_k\geq 0,\quad \sum_k P_k =1.
\end{equation}
In view of linearity property of tomographic map, one has the definition of
separability in terms of the state tomogram, i.e., the state is
separable, if
\begin{equation}\label{eq.R21}
\fl w(X_1,\mu_1,\nu_1,X_2,\mu_2,\nu_2)=\sum_k \,P_k\,
w_k^{(1)}(X_1,\mu_1,\nu_1)\,w_k^{(2)}(X_2,\mu_2,\nu_2).
\end{equation}
Tomogram is the joint probability-distribution function of two
random variables $X_1,X_2\,\in \mbox{R}$. Thus, the condition
of the state separability is formulated as the above property
(\ref{eq.R21}) of the joint probability distribution.

If tomogram cannot be written as convex sum (\ref{eq.R21}), the
state is entangled. The separability condition can be reformulated,
in view of the standard characteristic function for the tomographic
probability distribution (\ref{eq.R21}). In fact, if the
characteristic function can be written as
\begin{equation}\label{eq.R22}
\fl\chi(k_1,\mu_1,\nu_1,k_2,\mu_2,\nu_2)=\sum_k \,P_k
\,\chi_k^{(1)}(k_1,\mu_1,\nu_1)\,\chi_k^{(2)}(k_2,\mu_2,\nu_2)
\end{equation}
the state is separable. Here $\chi_k^{(1)}(k_1,\mu_1,\nu_1)$ and
$\chi_k^{(2)}(k_2,\mu_2,\nu_2)$ are the characteristic functions for
tomographic probabilities $w_k^{(1)}(X_1,\mu_1,\nu_1)$ and
$w_k^{(2)}(X_2,\mu_2,\nu_2)$, respectively.

An analogous definition of the separability and entanglement of the density
state $\widehat\rho(1,2)$ can be formulated in the diagonal representation.

Thus the state is separable, if the function which is symbol of the density
state in the diagonal representation can be written as
\begin{equation}\label{eq.R23}
\phi(z_1,z_2)=\sum_k \,P_k\,\phi_k^{(1)}(z_1)\,\phi_k^{(2)}(z_2).
\end{equation}

Thus we formulated the problem of separability and entanglement in
the diagonal representation of the density state $\widehat\rho(1,2)$. One can
easily extend the definition of separable and entangled states to
multipartite systems in both the tomographic and diagonal
representations of density states.

\section{Conclusions}
To conclude, we resume the main results of this work.

We reviewed the diagonal and probability representations of quantum states
using the standard star-product scheme. We found mutual relations of the
weight function of the diagonal representation and the tomographic-probability
distribution associated with the quantum state. We obtained the kernel of
star-product of operator symbols in the diagonal representation. The duality
relation between the diagonal representation of the weight function and
the Husimi--Kano function was obtained in the explicit form. The superposition
rule was formulated in both the diagonal representation and probability
representation of the density states. The problem of separability and
entanglement was formulated in both the diagonal and probability representations.

\section*{Acknowledgments}

V.I.M. thanks the Russian Foundation for Basic Research
for partial support under Projects Nos.~07-02-00598 and 08-02-90300
and the Organizers of the XV Central European Workshop on
Quantum Optics (Belgrade, Serbia, 30 May -- 3 June 2008)
for kind hospitality.

\end{document}